\documentclass[12pt]{article}
\usepackage[a4paper,margin=1in]{geometry}
\usepackage{amsmath,amssymb,graphicx,bm}
\usepackage{authblk}
\usepackage{setspace}
\usepackage{tikz}
\usepackage{float}

\title {Stiffness induced structures and morphological transitions in semiflexible polymers}

\author {Biman Bagchi}
\affil{Solid State and Structural Chemistry Unit, Indian Institute of Science, Bengaluru 560012, India}

\date{}

\begin{document}
\maketitle

%\begin{abstract}

\begin{abstract}
Semiflexible polymers in poor solvents exhibit a rich variety of collapsed
morphologies, including globules, toroids, and rodlike bundles, arising from the
competition between attractive interactions and chain stiffness. Computer
simulations and experiments on stiff and conjugated polymers have revealed
complex morphological crossovers, yet a unified theoretical description remains
incomplete. Here we develop a coarse-grained, field-theoretic free-energy
framework for linear polymers with variable stiffness that captures these
morphologies and their transitions within a common description. The theory is
built on three key ingredients: a density field describing monomer attraction
and excluded-volume effects, a nematic order parameter accounting for
orientational ordering in dense regions, and the bending rigidity of a
worm-like chain. Using simple variational ansatzes for competing morphologies,
we derive analytic expressions for their free energies and identify the
boundaries separating coil, globule, toroidal, and rodlike conformational
regimes as functions of the reduced attraction strength and the effective
persistence length. The resulting phase-diagram topology provides a
transparent free-energy-based framework for interpreting morphology diagrams
observed in simulations and experiments on semiflexible polymers in poor
solvents. We find the possibility of the existence of a triple point involving globules, rods and toroids.
\end{abstract}

%\end{abstract}

%=====================================
\section{Introduction}
%====================
Conformational transitions of single polymer chains in solution—such as the
change from an expanded coil to a compact globule—have been studied for many
decades and form one of the foundations of polymer physics.
The early works of Flory and de Gennes\cite{Flory1953,DeGennes1979} established
how a polymer in a good solvent behaves as a swollen coil, and how it collapses
into a compact state when the solvent quality worsens.
For flexible chains, this classical coil--globule transition is now well
understood as a balance between two competing effects: the loss of
conformational entropy upon chain compaction and the gain in attractive
interaction energy as monomers come closer together.\cite{GrosbergKhokhlov1994,Lifshitz1978}
Within this framework, the transition is often described at a mean-field level,
with thermal fluctuations primarily influencing the sharpness of the crossover
rather than its overall location.
Nevertheless, even in flexible polymers, fluctuations play an important role in
smoothing the transition and limiting the applicability of sharp phase
boundaries for finite chains.

The motivation for the present theoretical work stems from several computer simulation and experimental 
studies in which transformations are observed from globules to toroids and rods. These structures occur 
in a single polymer chain, and in polymers classified as sem-flexible chains. 

In fact, when the polymer is semiflexible, with a finite and tunable stiffness, the
phenomenology becomes considerably richer.
A stiff chain does not simply collapse in a poor solvent into a uniform
spherical globule; instead, it can adopt a variety of ordered morphologies,
including rods, racquet-like conformations, and toroids.
These structures arise because bending a semiflexible chain incurs an energetic
penalty, forcing the polymer to seek compact configurations that minimize
curvature while still benefiting from attractive interactions.
As a result, the competition between attraction and stiffness reshapes the
free-energy landscape and introduces new morphological regimes. [5-13]
Thermal fluctuations remain important in this context: they influence the
stability of competing morphologies, soften the boundaries between them, and
often render the transitions between different collapsed states as broad
crossovers rather than sharp phase changes. Needless to point out that fluctuarions add entropy to
a conformational state.

The central role of stiffness in governing such behavior was first identified in
early theoretical studies of semiflexible polymers\cite{KhokhlovSemenov1981,Odijk1986}
and later became especially prominent in the context of DNA condensation and
related biopolymers. [5-8]
%\cite{Bloomfield1997}

Experiments on DNA provide striking evidence of such ordered collapsed states.
In solutions containing multivalent counterions, DNA suddenly condenses into toroids or short rods with well-defined sizes and internal packing distances.\cite{Bloomfield1997,HudDowning2001,Pelta1996}
These observations stimulated major theoretical and computational efforts to understand why certain shapes are favored.
Both theory and simulations now show that the outcome of the collapse is governed by a competition between bending stiffness, surface tension, and self-attraction among monomers:
toroids minimize bending energy while keeping a compact shape, whereas rods minimize surface area but require sharper bends at the ends.\cite{UbbinkOdijk1999,Montesi2004,SolisOlvera1999}
Computational studies often employ Brownian and Langevin dynamics simulations of semi-flexible Lennard--Jones polymers. Such a study was carried out  by Srinivas and Bagchi who demonstrated existence of a sequence of conformations---approximately Gaussian coils, compact globules, rods, and toroids---upon tuning the LJ interaction strength and chain stiffness.\cite{SrinivasBagchi2002} In those works, the focus was on dynamical pathways and on detection of structural states using FRET-type probes.\cite{SrinivasBagchi2002,EnergyTransfer2002} Here we ask: can one construct a relatively simple field-theoretic free-energy description that (i) contains the same physical ingredients (variable chain stiffness and LJ interactions among the monomers) and (ii) obtain analytic expressions for the crossover values of stiffness and attraction at which the dominant morphology changes?

Bagchi, Rossky,Barbara et al. demonstrated in a \textit{Nature (2000} study that stiff conjugated polymers containing sparse chemical defects can undergo a spontaneous collapse from extended chains into highly ordered cylindrical conformations. \cite{Nature2000}  Using single-molecule fluorescence spectroscopy and theoretical modeling, they showed that even a small density of defects creates localized kinks that nucleate collapse and promote dense, nematically aligned packing along the chain backbone.  This work provided one of the earliest direct experimental demonstrations that semiflexible polymers can form ordered, rodlike morphologies through a defect-induced mechanism closely related to the bending--surface tension balance that governs toroid and rod formation in polymer condensation.

In this paper we provide such a description in a minimal Landau--de Gennes spirit. We treat the dense polymer as a ``nematic melt'' of segments with surface tension against the surrounding solvent, and we represent the conformations of interest (globule, toroid, rod) by simple geometrical ansatzes. By minimizing the free energy with respect to the relevant shape parameters we obtain scaling forms and crossover conditions. Although the resulting crossover lines are not universal numbers, they organize the phase behavior of LJ chains and connect naturally to experiments on DNA and other semiflexible biopolymers.

Many of the details of the derivations and background material are contained in the Appendices. In particular, we have discussed in detail  

\section{Semiflexible LJ chain: coarse-grained model}

We consider a linear chain of $N$ beads of diameter $a$, connected by harmonic
bonds of equilibrium length $b \approx a$. The total microscopic energy is
\begin{equation}
	U = U_{\mathrm{bond}} + U_{\mathrm{bend}} + U_{\mathrm{LJ}}.
\end{equation}
Bonded neighbors interact via
\begin{equation}
	U_{\mathrm{bond}} = \frac{k_b}{2} \sum_{i=1}^{N-1}
	\left( |\mathbf{r}_{i+1}-\mathbf{r}_i| - b \right)^2,
\end{equation}
while bending stiffness is introduced through
\begin{equation}
	U_{\mathrm{bend}} = \kappa \sum_{i=1}^{N-2} \left( 1 - \cos \theta_i \right),
\end{equation}
where $\theta_i$ is the angle between successive bond vectors, and $\kappa$ is
related to the persistence length $L_p$ by
\begin{equation}
	L_p = \frac{\kappa b}{k_B T}.
\end{equation}
Non-bonded beads interact through a Lennard--Jones potential
\begin{equation}
	U_{\mathrm{LJ}} = \sum_{i<j} 4\epsilon \left[
	\left( \frac{a}{r_{ij}} \right)^{12} -
	\left( \frac{a}{r_{ij}} \right)^{6} \right],
\end{equation}
with depth $\epsilon$ setting the solvent quality. Low $\epsilon$ corresponds to
good solvent (repulsive effective excluded volume), while increasing $\epsilon$
drives collapse and ordering.\cite{GrosbergKhokhlov1994}
%
% [Added A]
This microscopic bead--spring model serves as a reference system whose
short-length-scale degrees of freedom can be systematically coarse-grained,
leading to effective interaction parameters and free-energy functionals that
govern polymer behavior on mesoscopic length scales.

In the flexible limit $L_p \lesssim b$ and for weak attraction
$\epsilon \ll k_B T$, the chain behaves as a Gaussian coil with radius of
gyration
\begin{equation}
	R_g^2 \approx \frac{Nb^2}{6},
	\qquad (L_p \lesssim b,\ \epsilon \ll k_B T).
\end{equation}
%
% [Added B]
In this regime bending rigidity plays a negligible role, and the statistics are
well described by classical random-walk arguments.

For large $L_p$ the chain approaches the wormlike-chain (WLC)
regime.\cite{Yamakawa1997} In the collapsed state the relevant length scales are
the contour length $L = Nb$, the persistence length $L_p$, and a condensation
length that characterizes the balance between surface tension and
attraction.\cite{KhokhlovSemenov1981,Montesi2004}
%
% [Added C]
This condensation length reflects the competition between interfacial free
energy and bending elasticity and sets the characteristic dimensions of the
collapsed morphologies discussed below.
%=========================================================
%        {Field-theoretic free energy}
%==========================================================
%
%=========================================================
\section{Field-theoretic free energy}
%=========================================================
We now develop a coarse-grained field-theoretic free-energy functional for a
single semiflexible polymer, expressed in terms of a local monomer density
field and an orientational order parameter that characterizes segment alignment
in dense regions. The functional is constructed to incorporate effective two-
and three-body interactions, bending-induced nematic ordering at high density,
and the energetic cost of spatial inhomogeneity and interfaces. This unified
framework provides a transparent way to connect microscopic interactions and
chain stiffness to the emergence of globules, toroids, and rod-like morphologies,
and forms the basis for analyzing both mean-field behavior and fluctuation-
induced corrections.

\subsection{Density and orientational fields}

We introduce a coarse-grained monomer density field $\phi(\mathbf{r})$, defined
as the local monomer volume fraction, and a nematic order-parameter tensor
$\mathbf{Q}(\mathbf{r})$ describing the local orientational ordering of polymer
segments. Here $\phi(\mathbf{r})$ is understood as a coarse-grained quantity,
averaged over a length scale large compared to the monomer size but small
compared to the overall dimensions of the polymer.

At the mean-field level, the coarse-grained free-energy functional is written as
\begin{equation}
	\mathcal{F}[\phi,\mathbf{Q}]
	=
	\int d^3 r \left\{
	f_{\mathrm{dens}}(\phi)
	+ f_{\mathrm{int}}(\phi)
	+ f_{\mathrm{nem}}(\phi,\mathbf{Q})
	+ \frac{\kappa_{\mathrm{el}}}{2}(\nabla \phi)^2
	\right\}.
\end{equation}

The first term $f_{\mathrm{dens}}(\phi)$ is the density contribution to the free
energy,
\begin{equation}
	f_{\mathrm{dens}}(\phi)
	=
	k_B T \left[
	\phi \ln \phi + (1-\phi)\ln(1-\phi)
	\right],
\end{equation}
which represents the local mixing entropy of polymer segments and solvent. For
simplicity we adopt a lattice-gas form, which correctly captures the entropy of
mixing at the coarse-grained level, enforces the constraint $0\le\phi\le 1$, and
provides a simple way to incorporate excluded-volume effects at high density
while remaining analytically tractable.

The interaction part of the density free energy is taken as
\begin{equation}
	f_{\mathrm{int}}(\phi)
	=
	\frac{v}{2}\phi^2
	+
	\frac{w}{3}\phi^3,
\end{equation}
which is the standard virial expansion used in Flory-type and field-theoretic
descriptions of polymers.\cite{GrosbergKhokhlov1994}

Here $v$ and $w$ are coarse-grained interaction parameters. The coefficient $v$
is the renormalized second virial coefficient, which measures the net two-body
interaction between monomers after integrating out solvent degrees of freedom.
For a generic pair potential $U(r)$ it is given by
\begin{equation}
	v(T,\epsilon)
	=
	\int d^3 r \,
	\Big[1-\exp\!\big(-\beta U(r)\big)\Big],
\end{equation}
where $\beta=(k_BT)^{-1}$. For Lennard--Jones interactions, $v$ decreases
monotonically with increasing attraction strength $\epsilon$ and changes sign at
the $\theta$ condition, $v=0$, separating good- and poor-solvent regimes. In this
sense, increasing the Lennard--Jones attraction strength effectively worsens the
solvent quality by driving the renormalized second virial coefficient toward
negative values.

Microscopically, $v$ arises from the balance between excluded-volume repulsion
and attractive interactions and depends on both temperature and the strength of
intermolecular attractions. In good solvents $v>0$, while in poor solvents $v<0$,
signaling an effective attraction between monomers. Equivalently, $v$ may be
viewed as proportional to $(1-2\chi)$ in the Flory--Huggins description, where
$\chi$ is the polymer--solvent interaction parameter. We note that $v$ is a
strong function of temperature, as in liquid-state theory.

The coefficient $w>0$ represents an effective three-body interaction that
stabilizes the dense phase when $v<0$. Physically, it encodes short-range
repulsion arising from packing constraints and prevents collapse to infinite
density. At the coarse-grained level, $w$ is treated as weakly dependent on
temperature and attraction strength and is assumed positive throughout.

Within the Landau--Flory framework, the effective solvent quality is controlled
by $v(\epsilon,T)$ rather than by temperature alone, with increasing Lennard--
Jones attraction playing a role analogous to lowering temperature in a
conventional $\theta$ transition.

We now turn to the description of the orientational free energy. In the dense
collapsed state, semiflexible polymer segments behave as locally aligned rods,
and their collective orientational ordering can be captured by a nematic order
parameter. This motivates a Landau--de~Gennes description of orientational
ordering in dense polymer regions.

The orientational free-energy density is written as
\begin{equation}
	f_{\mathrm{nem}}(\phi,\mathbf{Q})
	=
	\frac{1}{2}A(\phi)\,\mathrm{Tr}\,\mathbf{Q}^2
	-
	\frac{1}{3}B\,\mathrm{Tr}\,\mathbf{Q}^3
	+
	\frac{1}{4}C\left(\mathrm{Tr}\,\mathbf{Q}^2\right)^2
	+
	\frac{K}{2}(\nabla\mathbf{Q})^2.
\end{equation}
The tensor $\mathbf{Q}(\mathbf{r})$ is symmetric and traceless and measures the
local degree of orientational alignment of polymer segments. In an isotropic
state $\mathbf{Q}=0$, while a nonzero $\mathbf{Q}$ indicates nematic-like
ordering.

The coefficients $A$, $B$, and $C$ are phenomenological Landau parameters that
control the onset and nature of orientational ordering. The quadratic
coefficient $A(\phi)$ acts as a ``mass'' term for the nematic order parameter and
determines the stability of the isotropic state. The cubic invariant
$\mathrm{Tr}\,\mathbf{Q}^3$ is symmetry allowed for nematic order and is
responsible for the generically first-order nature of the isotropic--nematic
transition in three dimensions.\cite{KhokhlovSemenov1981,Odijk1986} The quartic
term with $C>0$ ensures thermodynamic stability at large order-parameter
amplitude.

The gradient coefficient $K$ penalizes spatial variations of the orientational
tensor and represents the elastic cost of distortions of the local director
field. In the present context it controls the energetic penalty associated with
orientational inhomogeneities near interfaces and curved morphologies, and thus
acts as an orientational contribution to the effective surface tension of
ordered domains. This elastic penalty is distinct from the bending rigidity
$\kappa$ that controls the persistence length of the polymer backbone.

The coupling between density and orientational order is introduced through a
density-dependent nematic mass,
\begin{equation}
	A(\phi)=a_0(\phi-\phi^{*}),
\end{equation}
where $\phi^{*}$ is a threshold monomer volume fraction above which orientational
ordering becomes favorable. This linear dependence on density is the lowest-
order form consistent with symmetry and captures the physical fact that
orientational order emerges only beyond a critical packing fraction; higher-
order density dependences are neglected for simplicity.

We now turn to the description of interfacial effects. Spatial variations of the
density field are penalized by the square-gradient term
\begin{equation}
	\frac{\kappa_{\mathrm{el}}}{2}(\nabla\phi)^2,
\end{equation}
which enforces a smooth interface between dense polymer regions and the
surrounding dilute solution. Minimization of the free energy leads to a diffuse
interface whose energetic cost per unit area may be interpreted as an effective
surface tension.

Within square-gradient (Cahn--Hilliard) theory, the surface tension scales as \cite{CahnHilliard1958}
\begin{equation}
	\gamma \sim \sqrt{\kappa_{\mathrm{el}}\,\Delta f},
\end{equation}
where $\Delta f$ is the bulk free-energy density difference between the dense and
dilute phases. This scaling follows from minimizing the square-gradient free
energy across a one-dimensional interface and plays a central role in
determining the relative stability of globules, toroids, and rods.

We now briefly preview the role of fluctuations. Because the nematic mass
$A(\phi)$ depends explicitly on the local density, fluctuations of
$\phi(\mathbf{r})$ couple directly to the nematic amplitude through the invariant
$\mathrm{Tr}\,\mathbf{Q}^2(\mathbf{r})$. Because density fluctuations in the dense
phase are governed by a quadratic free energy, they can be integrated out
analytically, generating an effective interaction among orientational degrees of
freedom and renormalizing interfacial properties. These fluctuation effects
shift morphology boundaries and are analyzed in detail in
Sec.~\ref{sec:fluctuations}.

We refer to standard papers and textbooks in this field for background and methodology,
while the analyses presented here are new.
\cite{CahnHilliard1958,RowlinsonWidom1982,ChaikinLubensky1995,deGennesProst1993,Brazovskii1975}

We have provided several details of Landau-de Gennes free energy functional 
and Ornstein-Zernike correlation function in the Appendices.
%
% ======================= ROLE OF FLUCTUATIONS =========================
% ==================================================================
%============================Fluctuations =========================
\section{Role of fluctuations: density--nematic coupling and shifts of morphology boundaries}
\label{sec:fluctuations}
%==================================================================

A central motivation for a field-theoretic formulation is that it allows one to
analyze how \emph{fluctuations} of the density field $\phi(\mathbf{r})$ and the
orientational order parameter $\mathbf{Q}(\mathbf{r})$ modify the mean-field
morphology diagram. In the present problem, fluctuations are not merely
quantitative corrections but play a qualitatively important role because the
tendency toward orientational ordering depends explicitly on the local density.

The key nontrivial ingredient is that the coefficient controlling the quadratic
nematic term depends on $\phi(\mathbf{r})$. In particular, the Landau--de~Gennes
free-energy density contains a contribution of the form
\begin{equation}
	f_{\mathrm{nem}}(\phi,\mathbf{Q})
	=
	\frac{1}{2}A(\phi)\,\mathrm{Tr}\,\mathbf{Q}^2
	+\cdots,
	\qquad
	A(\phi)=a_0(\phi-\phi^{*}),
\end{equation}
where the dots denote higher-order and gradient terms in $\mathbf{Q}$. Local
densification ($\phi>\phi^{*}$) makes $A(\phi)$ negative and directly enhances
the tendency toward nematic ordering, while local density depletion suppresses
it. As a result, density fluctuations act as a fluctuating \emph{local control
	parameter} for orientational order. This mechanism implies that regions of
enhanced density naturally nucleate and stabilize nematic alignment.%------------------------------------------------------------------
\subsection{Gaussian density fluctuations and an induced interaction among nematic amplitudes}
%------------------------------------------------------------------

We consider fluctuations about a uniform dense reference state inside a
condensed polymer domain,
\begin{equation}
	\phi(\mathbf{r})=\phi_0+\delta\phi(\mathbf{r}).
\end{equation}
Expanding the density-dependent part of the free energy about $\phi_0$ to
quadratic order yields the Gaussian functional
\begin{equation}
	\mathcal{F}_{\phi}^{(2)}
	=
	\frac{1}{2}\int d^3 r\,
	\Big[
	r_{\phi}\,(\delta\phi)^2
	+
	\kappa_{\mathrm{el}}(\nabla\delta\phi)^2
	\Big],
	\qquad
	r_{\phi}\equiv
	\left.
	\frac{d^2}{d\phi^2}
	\big(
	f_{\mathrm{dens}}+f_{\mathrm{int}}
	\big)
	\right|_{\phi_0}.
\end{equation}
Here $r_\phi$ is the \emph{inverse compressibility} of the dense phase. A small
$r_\phi$ corresponds to a soft, highly compressible dense region with large
density fluctuations, whereas a large $r_\phi$ implies a stiff density field
and suppressed fluctuations.

The density dependence of $A(\phi)$ generates a term linear in $\delta\phi$,
\begin{equation}
	\mathcal{F}_{\phi Q}
	=
	\frac{a_0}{2}\int d^3 r\,
	\delta\phi(\mathbf{r})\,\mathrm{Tr}\,\mathbf{Q}^2(\mathbf{r})
	+
	\frac{a_0}{2}(\phi_0-\phi^{*})
	\int d^3 r\,\mathrm{Tr}\,\mathbf{Q}^2(\mathbf{r}).
\end{equation}
The second term represents the mean-field shift of the nematic ``mass'' at the
reference density $\phi_0$. The first term is the essential fluctuation coupling:
positive density fluctuations locally reduce $A(\phi)$ and therefore promote
nematic ordering. In this sense, density fluctuations actively enhance
orientational order rather than merely correlating with it.

Because $\delta\phi$ is Gaussian at this level, it can be integrated out
analytically. Completing the square shows that eliminating $\delta\phi$
\emph{always lowers} the free energy in regions where
$\mathrm{Tr}\,\mathbf{Q}^2$ is large, leading to an \emph{attractive} effective
interaction among nematic amplitude fluctuations,
\begin{equation}
	\Delta\mathcal{F}_{\mathrm{eff}}[\mathbf{Q}]
	=
	-\frac{a_0^2}{8}
	\int d^3 r\,d^3 r'\;
	\mathrm{Tr}\,\mathbf{Q}^2(\mathbf{r})\,
	G_{\phi}(\mathbf{r}-\mathbf{r}')\,
	\mathrm{Tr}\,\mathbf{Q}^2(\mathbf{r}').
\end{equation}
The density correlator is
\begin{equation}
	G_{\phi}(\mathbf{q})
	=
	\frac{1}{r_{\phi}+\kappa_{\mathrm{el}}q^2},
	\qquad
	\xi_{\phi}=\sqrt{\kappa_{\mathrm{el}}/r_{\phi}},
\end{equation}
so that $\xi_\phi$ defines the density correlation length inside the dense
phase. This length sets the spatial range over which density fluctuations
mediate effective interactions between nematic amplitudes.

Here $r_{\phi}$ ensures local thermodynamic stability of the dense phase
($r_\phi>0$), while its magnitude controls the compressibility of the
polymer-rich region. Together with $\kappa_{\mathrm{el}}$, it determines
$\xi_{\phi}$ and therefore the spatial extent over which density fluctuations
influence orientational ordering.

For slowly varying $\mathbf{Q}$ fields on length scales much larger than
$\xi_{\phi}$, the induced interaction becomes effectively local. In this limit,
density fluctuations renormalize the quartic coefficient in the nematic free
energy,
\begin{equation}
	\frac{C}{4}\big(\mathrm{Tr}\,\mathbf{Q}^2\big)^2
	\;\longrightarrow\;
	\frac{C_{\mathrm{eff}}}{4}\big(\mathrm{Tr}\,\mathbf{Q}^2\big)^2,
	\qquad
	C_{\mathrm{eff}}\simeq C-\frac{a_0^2}{2r_{\phi}}.
\end{equation}
Thus density fluctuations deepen the ordered minimum and enhance the tendency
toward nematic ordering inside the dense polymer phase. When
$a_0^2/(2r_\phi)$ is appreciable, the effective quartic stiffness is reduced,
increasing the susceptibility of the ordering transition to fluctuation
effects.
%------------------------------------------------------------------
\subsection{Implications for morphology boundaries: why globule--toroid shifts}
%------------------------------------------------------------------

The globule--toroid crossover compares two \emph{dense} morphologies, so the bulk
term $N f_{\mathrm{bulk}}(\phi_0)$ largely cancels. The competition is therefore
controlled primarily by surface tension and bending elasticity, both of which
are sensitive to density--nematic coupling.

\paragraph{(i) Facilitation of orientational order.}
Local positive density fluctuations $\delta\phi>0$ make $A(\phi)$ more negative
and directly favor nematic ordering. This mechanism preferentially stabilizes
ordered dense morphologies (toroids and rods) relative to an isotropic globule.
The effect is strongest near the onset of orientational order, where the nematic
susceptibility is large and $\mathrm{Tr}\,\mathbf{Q}^2$ responds strongly to
small changes in density.

\paragraph{(ii) Renormalization of the effective surface tension.}
The interface between dense and dilute regions is described at the coarse-
grained level by $\kappa_{\mathrm{el}}(\nabla\phi)^2/2$. Density fluctuations
broaden the interface and renormalize its free energy. Moreover, because
nematic order is favored in the dense interior but suppressed in the interfacial
region, orientational degrees of freedom contribute an additional interfacial
penalty. At the level of shape free energies, these effects can be summarized by
a replacement $\gamma\rightarrow\gamma_{\mathrm{eff}}$.

Since the globule surface cost scales as
$F_{\mathrm{glob}}'\sim \gamma\,a^2 N^{2/3}$ whereas the minimized toroid cost
scales as $F_{\mathrm{tor}}'\sim (k_BT)^{2/5}L_p^{2/5}\gamma^{4/5}\cdots$, even a
modest renormalization of $\gamma$ can shift the globule--toroid boundary. In
particular, fluctuation-induced softening of the interfacial penalty favors the
toroid relative to the spherical globule because the latter depends more strongly
on $\gamma$.

Taken together, density fluctuations coupled through
$A(\phi)=a_0(\phi-\phi^{*})$ generically shift the globule--toroid crossover so
that toroids become stable at slightly smaller stiffness $L_p^{*}$ (or slightly
weaker attraction $\epsilon^{*}$) than predicted by a strict mean-field
treatment. The magnitude of this shift is controlled by the compressibility
parameter $r_{\phi}$ and the density correlation length $\xi_{\phi}$ in the
dense phase, and is expected to be most pronounced near the onset of nematic
ordering where the density-enhanced nematic susceptibility is largest.

%==================================================================
%===================================Old Section 5  deleted. This is nw. ========================
% 
%==================================Coil & Globule ========================================================

\section{Coil and globule free energies}

Here we apply the formalism developed above to derive expressions for the free
energies of the coil and globule states. In later sections we obtain analogous
expressions for the ordered morphologies (rod and toroid). The present section
highlights how the competition between entropy and interaction energy favors
distinct polymer conformations.

%------------------------------
\subsection{Gaussian coil}

The Flory free energy for a flexible polymer coil of size $R$ is
given by\cite{Flory1953,DeGennes1979}
\begin{equation}
	\frac{F_{\mathrm{coil}}(R)}{k_B T}
	=
	\frac{3 R^2}{2 N b^2}
	+
	\frac{v N^2}{R^3}
	+
	\frac{w N^3}{R^6}.
\end{equation}
Minimization with respect to $R$ yields the equilibrium coil size and the
corresponding free energy. In a good solvent ($v>0$) and for chains that are not
too long, the three-body term can be neglected, giving
\begin{equation}
	R_{\mathrm{coil}}
	\sim
	b N^{3/5}
	\left( \frac{v}{b^3} \right)^{1/5},
\end{equation}
and
\begin{equation}
	F_{\mathrm{coil}}
	\sim
	\frac{5}{2}\,k_B T\,N^{1/5}
	\left(\frac{v}{b^3}\right)^{2/5}.
\end{equation}

As the solvent quality worsens, the effective excluded-volume coefficient $v$
decreases and passes through zero at the $\theta$ point. For $v<0$,
monomer--monomer attractions dominate, rendering the swollen coil
thermodynamically unstable. The three-body repulsion, represented by the
$wN^3/R^6$ term, prevents unphysical collapse and stabilizes a dense globule of
finite monomer density.

%------------------------------
\subsection{Compact globule}

In the collapsed state we assume a uniform dense globule of volume $V$ and radius
$R_{\mathrm{glob}}$ containing all $N$ monomers at a fixed density $\phi_0$. One
then has
\begin{equation}
	V = \frac{4\pi}{3} R_{\mathrm{glob}}^3 = \frac{N a^3}{\phi_0}.
\end{equation}
The bulk free energy is
\begin{equation}
	F_{\mathrm{bulk,glob}} = N f_{\mathrm{bulk}}(\phi_0),
\end{equation}
where $f_{\mathrm{bulk}}(\phi_0)$ is the bulk free energy per monomer in the dense
phase, obtained by evaluating $f_{\mathrm{dens}}+f_{\mathrm{int}}$ at $\phi_0$.

The surface contribution arises from the interface between the dense globule and
the surrounding dilute solution and is given by
\begin{equation}
	F_{\mathrm{surf,glob}}
	=
	4 \pi R_{\mathrm{glob}}^2 \gamma
	=
	4\pi \gamma
	\left(
	\frac{3N a^3}{4\pi \phi_0}
	\right)^{2/3}.
\end{equation}
Thus, for large $N$,
\begin{equation}
	F_{\mathrm{glob}}
	\approx
	N f_{\mathrm{bulk}}(\phi_0)
	+
	\tilde{A}_{\mathrm{glob}}\,\gamma a^2 N^{2/3},
\end{equation}
with the geometric prefactor
\begin{equation}
	\tilde{A}_{\mathrm{glob}}
	=
	4\pi
	\left(
	\frac{3}{4\pi \phi_0}
	\right)^{2/3}.
\end{equation}

The coil--globule crossover can be estimated by equating $F_{\mathrm{coil}}$ and
$F_{\mathrm{glob}}$ at a given chain length $N$, yielding a relation between the
effective excluded-volume parameter $v$ (or equivalently the attraction strength
$\epsilon$) and $N$,
\begin{equation}
	N f_{\mathrm{bulk}}(\phi_0)
	+
	\tilde{A}_{\mathrm{glob}}\,\gamma a^2 N^{2/3}
	\approx
	k_B T\,N^{1/5}
	\left(\frac{v}{b^3}\right)^{2/5}.
\end{equation}

Solving this relation for $v$ defines the effective coil--globule transition line
in the $(N,v)$ plane. For very long chains the surface contribution becomes
relatively less important, and the transition is controlled primarily by the
sign change of the bulk free-energy density.
%
%=====================Toroid versus Rods =========================
%%=============================================
\section{Nematic toroid and rod: bending versus surface tension}
%=============================================

In the dense regime, semiflexibility and orientational ordering become central in
determining polymer morphology.\cite{KhokhlovSemenov1981,Odijk1986,UbbinkOdijk1999}
Previous approaches have typically estimated rodlike and toroidal conformations
by balancing bending elasticity against surface tension in a largely geometric
manner. Here we derive the free energies of nematic toroids and rods within the
\emph{same coarse-grained field-theoretic framework} developed above. In this
unified description, surface tension, bending rigidity, and orientational order
all emerge from a common free-energy functional, allowing a transparent and
self-consistent comparison of competing dense morphologies and preparing the
ground for constructing morphology phase boundaries.

The dominant competition controlling the shapes of collapsed states is between
bending energy and surface tension. Bending costs are reduced by stiffness, while
attractive interactions favor compactness and promote nematic alignment in dense
regions. We now derive physically motivated expressions for toroidal and rodlike
conformations.

%==========================
\subsection{Toroid}
%==========================

A toroid is a particularly interesting geometry in which curvature is distributed
uniformly along the polymer backbone. We model the toroid as a circular tube of
minor radius $r$ whose center traces a circle of major radius $R$. The total
contour length $L=Nb$ is packed into a toroidal volume of cross-sectional area
$\pi r^2$, giving the volume constraint
\begin{equation}
	2\pi R \cdot \pi r^2 = \frac{N a^3}{\phi_0}.
\end{equation}

The dominant curvature of the polymer backbone arises from bending around the
major radius $R$, while curvature associated with the minor radius is neglected
at this level. The bending energy is therefore estimated as\cite{Odijk1986,UbbinkOdijk1999}
\begin{equation}
	F_{\mathrm{bend,tor}}
	\approx
	\frac{\kappa}{2} \int_0^L ds\, \frac{1}{R^2}
	=
	\frac{\kappa L}{2 R^2}.
\end{equation}

The surface area of the torus is
\begin{equation}
	A_{\mathrm{tor}} = 4\pi^2 R r,
\end{equation}
so the surface free energy is
\begin{equation}
	F_{\mathrm{surf,tor}} = 4 \pi^2 \gamma R r.
\end{equation}

Eliminating $r$ using the volume constraint,
\begin{equation}
	r^2 = \frac{N a^3}{2\pi^2 \phi_0 R},
\end{equation}
yields
\begin{equation}
	F_{\mathrm{surf,tor}}
	=
	4\pi^2 \gamma R
	\sqrt{\frac{N a^3}{2\pi^2 \phi_0 R}}
	=
	\tilde{B}_{\mathrm{tor}}\,
	\gamma\, a^{3/2} N^{1/2} R^{1/2},
\end{equation}
with $\tilde{B}_{\mathrm{tor}} = 4\pi^2 (2\pi^2\phi_0)^{-1/2}$.

Subtracting the common dense bulk contribution $N f_{\mathrm{bulk}}(\phi_0)$,
which is identical for all collapsed morphologies, the reduced toroidal free
energy becomes
\begin{equation}
	F_{\mathrm{tor}}'(R)
	=
	\frac{\kappa L}{2 R^2}
	+
	\tilde{B}_{\mathrm{tor}} \gamma a^{3/2} N^{1/2} R^{1/2}.
\end{equation}

Minimization with respect to $R$ gives
\begin{equation}
	-\kappa L R^{-3}
	+
	\frac{1}{2}\tilde{B}_{\mathrm{tor}} \gamma a^{3/2} N^{1/2} R^{-1/2}
	=0,
\end{equation}
or
\begin{equation}
	R_{\mathrm{tor}}
	\sim
	\left(
	\frac{\kappa L}{\gamma a^{3/2} N^{1/2}}
	\right)^{2/5}.
\end{equation}

Substitution into $F_{\mathrm{tor}}'$ yields the minimized toroidal free energy,
\begin{equation}
	F_{\mathrm{tor}}'
	\sim
	(k_B T)^{2/5}
	L_p^{2/5}
	\gamma^{4/5}
	a^{6/5} b^{1/5}
	N^{2/5},
\end{equation}
which explicitly displays the combined dependence on stiffness and surface
tension.

%==========================
\subsection{Rod}
%==========================

We next consider a rodlike bundle of length $H$ and radius $R_{\mathrm{rod}}$.
The volume constraint is
\begin{equation}
	\pi R_{\mathrm{rod}}^2 H = \frac{N a^3}{\phi_0}.
\end{equation}

In a rod, polymer segments are predominantly aligned along the long axis, and
bending occurs mainly near the end caps. A simple estimate treats each cap as a
semicircle of radius $R_{\mathrm{rod}}$, giving a bending contribution
\begin{equation}
	F_{\mathrm{bend,rod}} \sim \kappa \frac{\pi}{R_{\mathrm{rod}}}.
\end{equation}

The surface area is
\begin{equation}
	A_{\mathrm{rod}} = 2\pi R_{\mathrm{rod}} H + 2\pi R_{\mathrm{rod}}^2,
\end{equation}
leading to a surface free energy
\begin{equation}
	F_{\mathrm{surf,rod}}
	=
	2\pi \gamma R_{\mathrm{rod}} H
	+
	2\pi \gamma R_{\mathrm{rod}}^2.
\end{equation}

Eliminating $H$ using the volume constraint,
\begin{equation}
	H = \frac{N a^3}{\pi \phi_0 R_{\mathrm{rod}}^2},
\end{equation}
one finds
\begin{equation}
	F_{\mathrm{surf,rod}}
	=
	\frac{2\gamma N a^3}{\phi_0 R_{\mathrm{rod}}}
	+
	2\pi \gamma R_{\mathrm{rod}}^2.
\end{equation}

Subtracting the bulk contribution, the reduced rod free energy is
\begin{equation}
	F_{\mathrm{rod}}'(R_{\mathrm{rod}})
	\sim
	\kappa \frac{\pi}{R_{\mathrm{rod}}}
	+
	\frac{2\gamma N a^3}{\phi_0 R_{\mathrm{rod}}}
	+
	2\pi \gamma R_{\mathrm{rod}}^2.
\end{equation}

For sufficiently large $N$, the surface term proportional to $N/R_{\mathrm{rod}}$
dominates over the bending contribution for realistic stiffness values, allowing
the approximation
\begin{equation}
	F_{\mathrm{rod}}'
	\approx
	\frac{2\gamma N a^3}{\phi_0 R_{\mathrm{rod}}}
	+
	2\pi \gamma R_{\mathrm{rod}}^2.
\end{equation}

Minimization yields
\begin{equation}
	R_{\mathrm{rod}}^3
	\sim
	\frac{N a^3}{2\pi \phi_0},
\end{equation}
and the minimized rod free energy
\begin{equation}
	F_{\mathrm{rod}}'
	\sim
	\gamma a^2 N^{2/3}.
\end{equation}

The rod free energy therefore scales with chain length in the same manner as the
surface term of a spherical globule, but with a different geometry-dependent
prefactor.
%===================================================
%==============Crossover Conditions : Section =========================
%======================================================
%
\section{Crossover conditions}

The free-energy expressions derived in the preceding sections provide a unified
framework for analyzing crossovers between distinct polymer morphologies.
Because four conformational states are involved—coil, globule, toroid, and rod—
multiple crossover lines arise, leading to a rich phase behavior in the space of
chain length, stiffness, and attractive interaction strength.
In this section we derive the conditions for these crossovers and elucidate their
physical origin. Numerical illustrations are presented subsequently.

%=====================================
\subsection{Coil--globule crossover}
%=====================================

The coil--globule crossover is obtained by comparing the Flory free energy of an
expanded coil with the free energy of a compact globule of fixed monomer volume
fraction $\phi_0$.

\paragraph{Coil free energy.}
For a flexible chain of size $R$, the Flory free energy reads
\begin{equation}
	\frac{F_{\mathrm{coil}}(R)}{k_B T}
	=
	\frac{3R^2}{2Nb^2}
	+ \frac{vN^2}{R^3},
\end{equation}
where $b$ is the Kuhn length and $v$ is the effective second virial coefficient.
Minimization with respect to $R$ yields the standard Flory estimate
\begin{equation}
	R_{\mathrm{coil}}
	\sim
	b N^{3/5} \left( \frac{v}{b^3} \right)^{1/5}.
\end{equation}
Substitution back into $F_{\mathrm{coil}}$ gives
\begin{equation}
	F_{\mathrm{coil}}
	\sim
	\frac{5}{2}\,k_B T\,N^{1/5} \left( \frac{v}{b^3}\right)^{2/5} .
\end{equation}
This free energy is positive and purely entropic in origin; it decreases as
$v \rightarrow 0^+$, approaching the $\theta$-point.

\paragraph{Globule free energy.}
In the collapsed state the polymer forms a dense phase of monomer volume fraction
$\phi_0$ with total volume $V = N a^3/\phi_0$. The globule radius is therefore
\begin{equation}
	R_{\mathrm{glob}}
	=
	\left(
	\frac{3Na^3}{4\pi\phi_0}
	\right)^{1/3}.
\end{equation}
The globule free energy consists of a bulk contribution and an interfacial term,
\begin{equation}
	F_{\mathrm{glob}}
	=
	N f_{\mathrm{bulk}}(\phi_0)
	+ 4\pi R_{\mathrm{glob}}^2 \gamma
	=
	N f_{\mathrm{bulk}}(\phi_0)
	+ \tilde{A}_{\mathrm{glob}}\,\gamma a^2 N^{2/3},
\end{equation}
where $\gamma$ is the surface tension and $\tilde{A}_{\mathrm{glob}}$ is a geometric
constant of order unity. Both contributions are positive: the bulk term reflects
the cost of forming a dense phase, while the surface term penalizes compactification.

\paragraph{Crossover condition.}
The coil--globule crossover is defined by the condition
$F_{\mathrm{coil}} = F_{\mathrm{glob}}$.
In the asymptotic large-$N$ limit the surface contribution becomes negligible
compared to the bulk term, and the crossover reduces to the classical Flory
condition $v=0$, corresponding to the $\theta$-point.
For finite chains, however, the surface term induces a shift in the effective
crossover value of $v$.

Equating the two free energies yields
\begin{equation}
N f_{\mathrm{bulk}}(\phi_0)
+ \tilde{A}_{\mathrm{glob}} \gamma a^2 N^{2/3}
\simeq
\frac{5}{2}\,k_B T\,N^{1/5}
\left( \frac{v}{b^3} \right)^{2/5}.
\end{equation}
Solving for $v$ gives the $N$-dependent finite-size crossover line,
\begin{equation}
v_{\mathrm{c}}(N)
\sim
b^3
\left[
\frac{f_{\mathrm{bulk}}(\phi_0)}{k_B T} N^{4/5}
+
\tilde{A}_{\mathrm{glob}}
\frac{\gamma a^2}{k_B T} N^{3/5}
\right]^{5/2}.
\end{equation}

\paragraph{Physical interpretation.}
For an infinite chain the surface contribution vanishes relative to the bulk
term, and $v_{\mathrm{c}}\to 0$, recovering the standard $\theta$-condition.
For finite chains, however, surface tension stabilizes the coil and shifts the
crossover to slightly positive values of $v$.
Strongly negative $v$ corresponds to genuinely poor-solvent conditions, where the
globule description is appropriate and the Flory coil free energy no longer applies.

%=====================================
\subsection{Globule--toroid crossover}
%=====================================

In the dense regime the relevant competition is between globular and toroidal
conformations. Since both share the same bulk free energy, the crossover is
determined by comparing the excess free energies,
\begin{equation}
F_{\mathrm{glob}}' \sim \tilde{A}_{\mathrm{glob}} \gamma a^2 N^{2/3},
\qquad
F_{\mathrm{tor}}' \sim 
(\kappa \gamma^2 a^3)^{2/5} L^{1/5} N^{1/5},
\end{equation}
where $\kappa$ is the bending modulus and $L=Nb$ is the contour length.
Equating these expressions gives
\begin{equation}
(\kappa \gamma^2 a^3)^{2/5} (Nb)^{1/5} N^{1/5}
\sim
\tilde{A}_{\mathrm{glob}} \gamma a^2 N^{2/3}.
\end{equation}
Solving for the persistence length $L_p=\kappa/(k_BT)$ yields
\begin{equation}
L_{p,\mathrm{c}}^{\mathrm{glob}\rightarrow\mathrm{tor}}
\sim
\tilde{A}_{\mathrm{glob}}^{5/2}
\gamma^{1/2} a^2 b^{-1/2}
N^{4/15}.
\end{equation}
Thus the stiffness required to stabilize a toroid relative to a globule increases
with chain length as $N^{4/15}$ and grows with increasing surface tension.

%=====================================
\subsection{Globule--rod and toroid--rod crossovers}
%=====================================

Comparing the excess free energies of rods and globules,
\begin{equation}
F_{\mathrm{rod}}' \sim c_{\mathrm{rod}} \gamma a^2 N^{2/3},\qquad
F_{\mathrm{glob}}' \sim \tilde{A}_{\mathrm{glob}} \gamma a^2 N^{2/3},
\end{equation}
shows that both scale identically with $N$ and differ only by geometric prefactors.
At this level of description, rods and globules are therefore nearly degenerate.
In practice, nematic ordering and end-cap bending effects favor toroidal
conformations for sufficiently stiff chains, while rods can be stabilized by
confinement or external alignment fields.

The more informative competition is between rods and toroids. Equating
$F_{\mathrm{rod}}'$ and $F_{\mathrm{tor}}'$ gives
\begin{equation}
(\kappa \gamma^2 a^3)^{2/5} (Nb)^{2/5}
\sim
c_{\mathrm{rod}} \gamma a^2 N^{2/3}.
\end{equation}
Solving for $\kappa$ yields
\begin{equation}
\kappa_{\mathrm{c}}^{\mathrm{rod}\leftrightarrow\mathrm{tor}}
\sim
k_B T\, c_{\mathrm{rod}}^{5/2}
\gamma^{-1/2} a^{1/2} b^{-1}
N^{4/15}.
\end{equation}
Both the globule--toroid and rod--toroid crossover lines therefore exhibit the same
characteristic $N^{4/15}$ scaling, differing only in their prefactors.

In the $(L_p,\epsilon)$ plane (with $\gamma$ an increasing function of the LJ
attraction strength $\epsilon$), these relations reproduce the qualitative phase
structure observed in simulations: a globular regime at small stiffness and
moderate attraction, a toroidal regime at large stiffness and strong attraction,
and an intermediate rod-like region.

%=====================================
%===================================================
\section{Phase diagram in dimensionless variables and comparison with simulations}
%==================================================
A central outcome of the field-theoretic formulation developed above is that the
competition between Lennard--Jones attraction, bending rigidity, and surface
tension can be expressed in terms of two natural dimensionless control parameters.
These parameters organize all observed polymer morphologies—coil, globule, rod,
and toroid—into distinct regions of a single schematic phase diagram.
In this section we construct this diagram and discuss how it rationalizes the
Brownian-dynamics simulations of Srinivas and Bagchi,\cite{SrinivasBagchi2002} as
well as earlier theoretical and experimental studies of semiflexible polymer
collapse.

%==================================================
\subsection{Dimensionless axes}
%==================================================

Following standard practice in polymer physics,\cite{Flory1953,DeGennes1979,GrosbergKhokhlov1994}
we introduce two dimensionless variables:

\begin{enumerate}
	\item \textbf{Reduced Lennard--Jones attraction strength}
	\begin{equation}
	\epsilon^{*} \equiv \frac{\epsilon}{k_B T}.
	\end{equation}
	Increasing $\epsilon^{*}$ corresponds to poorer solvent conditions and stronger
	monomer attraction. The effective second virial coefficient $v(\epsilon^{*})$
	decreases monotonically with $\epsilon^{*}$ and crosses zero at the $\theta$-point,
	beyond which dense globules are thermodynamically favored.\cite{Lifshitz1978}
	
	\item \textbf{Reduced persistence length (stiffness)}
	\begin{equation}
	L_p^{*} \equiv \frac{L_p}{b}
	= \frac{\kappa}{k_B T\, b}.
	\end{equation}
	Here $L_p$ is the persistence length of the wormlike chain, $b$ is the bond length,
	and $\kappa$ is the bending modulus. This parameter distinguishes flexible chains
	($L_p^{*}\ll1$) from semiflexible and rodlike chains ($L_p^{*}\gg1$).\cite{Odijk1986,Yamakawa1997,KhokhlovSemenov1981}
\end{enumerate}

The $(\epsilon^{*},L_p^{*})$ plane therefore provides a compact and physically
transparent representation of the balance between entropic elasticity, excluded
volume, attractive interactions, and bending rigidity.

%==================================================
\subsection{Structure of the phase diagram}
%==================================================
Three crossover lines determine the qualitative morphology.

\paragraph{(i) Coil--globule boundary.}

Within Flory theory the coil--globule crossover is controlled primarily by the
sign of the effective second virial coefficient,
\begin{equation}
v(\epsilon^{*}) \approx 0,
\end{equation}
corresponding to a $\theta$-like transition.
For sufficiently long chains this condition is essentially independent of
stiffness, and the coil--globule boundary appears as an approximately vertical
line in the $(\epsilon^{*},L_p^{*})$ plane.
Finite-size effects introduce weak corrections, but flexible and semiflexible
chains collapse at comparable values of $\epsilon^{*}$.

\paragraph{(ii) Globule--nematic (ordered) boundary.}

Within the dense phase, orientational ordering arises when bending rigidity
overcomes orientational entropy.
From the Landau analysis of nematic ordering inside the globule (Eq.~50), the
onset condition may be written as
\begin{equation}
L_p > L_p^{\ast}(\epsilon^{*})
= L_p^0(T)
+ \frac{2\lambda}{a_Q}\,\phi_0(\epsilon^{*}),
\end{equation}
where $\phi_0(\epsilon^{*})$ is the equilibrium monomer volume fraction in the
dense phase.
Because $\phi_0$ increases monotonically with attraction strength, this boundary
slopes upward with increasing $\epsilon^{*}$: denser globules stabilize nematic
order at lower stiffness.
This line separates isotropic globules from orientationally ordered condensed
states (rods and toroids).

\paragraph{(iii) Toroid--rod boundary.}

Within the ordered regime, the relevant competition is between bending energy and
surface energy.
Equating the minimized toroidal and rod free energies yields the scaling
\begin{equation}
L_p^{(\mathrm{tor\!-\!rod})}(\epsilon^{*})
\propto 
\gamma(\epsilon^{*})^{-1/2}\, N^{4/15},
\end{equation}
where $\gamma(\epsilon^{*})$ is the surface tension of the dense phase.
Since $\gamma$ increases with $\epsilon^{*}$,\cite{UbbinkOdijk1999} this boundary
slopes downward: stronger attractions favor toroids over rods because surface
tension penalizes rod end-caps more severely than the smoothly curved toroidal
interface.

%==================================================
\subsection{Qualitative topology of the phase diagram}
%==================================================
The combination of these three boundaries leads to the universal topology shown
schematically in Fig.~\ref{fig:phase-diagram}:

\begin{itemize}
	\item \textbf{Low $\epsilon^{*}$, low $L_p^{*}$:} expanded Gaussian coils.
	\item \textbf{Intermediate $\epsilon^{*}$, low $L_p^{*}$:} isotropic globules.
	\item \textbf{Intermediate $\epsilon^{*}$, high $L_p^{*}$:} rodlike conformations.
	\item \textbf{High $\epsilon^{*}$, moderate $L_p^{*}$:} toroids stabilized by
	surface tension and moderate stiffness.
\end{itemize}

This topology is qualitatively identical to early theoretical predictions for
semiflexible polymer condensation,\cite{KhokhlovSemenov1981,Odijk1986} and to the
experimentally observed morphologies of DNA under condensing conditions.\cite{Bloomfield1997,HudDowning2001,Pelta1996,UbbinkOdijk1999}

%==================================================
\subsection{Comparison with the Srinivas--Bagchi simulations}
%==================================================

Srinivas and Bagchi performed extensive Brownian-dynamics simulations of a single
semiflexible Lennard--Jones chain with tunable attraction strength and bending
rigidity.\cite{SrinivasBagchi2002}
Their observed morphologies map naturally onto the regions predicted by the
theory:

\begin{itemize}
	\item \textbf{Coils} for $\epsilon^{*}\lesssim 1.0$, largely independent of stiffness.
	\item \textbf{Globules} for $\epsilon^{*}\gtrsim 1.1$ and $L_p^{*}\lesssim 2$.
	\item \textbf{Toroids} for $\epsilon^{*}\approx 1.3$--$1.5$ and moderate stiffness
	($L_p^{*}\sim 3$--$8$).
	\item \textbf{Rods} for sufficiently stiff chains ($L_p^{*}\gtrsim 8$), across a
	wide range of attraction strengths.
\end{itemize}

These simulation results are overlaid schematically in
Fig.~\ref{fig:phase-diagram}.
Although the theoretical boundaries shown are not intended to be quantitatively
precise, the agreement in morphology sequence and phase topology is striking.
The essential features follow directly from the minimal free-energy framework
developed here, without adjustable parameters beyond those entering the
coarse-grained description.

\subsection{Schematic phase diagram}
The schematic phase diagram is shown in Fig.~\ref{fig:phase-diagram}.
The phase diagram may be visualized using the following figure
(Figure 1).
%
% ---- INSERT TIKZ CODE PROVIDED EARLIER HERE ----=====
%======================= Figure Below ======================

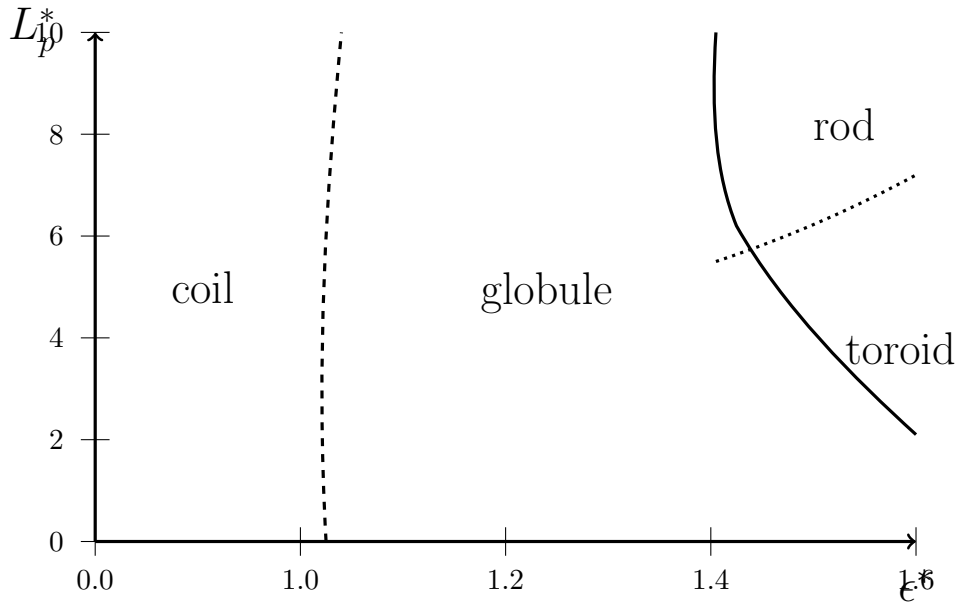
\begin{figure}[H]
	\centering
	\begin{tikzpicture}[scale=1.35]
	
	%========================
	% Axes
	%========================
	\draw[->, very thick] (0,0) -- (8,0)
	node[below, yshift=-6pt] {\Large $\epsilon^{*}$};
	
	\draw[->, very thick] (0,0) -- (0,5)
	node[left, xshift=-10pt] {\Large $L_p^{*}$};
	
	%========================
	% X-axis ticks
	%========================
	\foreach \x/\label in {0/0.0, 2/1.0, 4/1.2, 6/1.4, 8/1.6}
	{
		\draw (\x,0.14) -- (\x,-0.14);
		\node[below] at (\x,-0.18) {\small \label};
	}
	
	%========================
	% Y-axis ticks
	%========================
	\foreach \y/\label in {0/0, 1/2, 2/4, 3/6, 4/8, 5/10}
	{
		\draw (0.14,\y) -- (-0.14,\y);
		\node[left] at (-0.20,\y) {\small \label};
	}
	
	%====================================================
	% PHASE BOUNDARIES (P--T style)
	%====================================================
	
	% Coil--globule boundary
	\draw[very thick, dashed]
	(2.25,0.0) .. controls (2.15,1.8) and (2.25,3.4) .. (2.40,5.0);
	
	% Globule--toroid boundary
	% Globule--toroid boundary:
	% toroids do NOT exist below L_p^* = L_min^*, so boundary turns right and ends.
	\def\Lmin{1.05}
	
	\draw[very thick]
	(6.05,5.00)
	.. controls (6.00,4.20) and (6.05,3.60)
	.. (6.25,3.10)
	.. controls (6.70,2.30) and (7.40,1.60)
	.. (8.00,\Lmin);

	%===========================
	% Toroid--rod boundary (lowered)
	\draw[very thick, dotted]
	(6.05,2.75) .. controls (6.80,3.00) and (7.45,3.30) .. (8.00,3.60);
	
	%====================================================
	% PHASE LABELS
	%====================================================
	
	\node at (1.05,2.50) {\Large coil};
	
	\node at (4.40,2.45) {\Large globule};
	
	\node at (7.85,1.90) {\Large toroid};
	
	\node at (7.30,4.10) {\Large rod};
	
	\end{tikzpicture}
	
	\caption{Schematic P--T--type phase diagram of a semiflexible polymer in the plane of
		reduced attraction strength $\epsilon^{*}=\epsilon/k_BT$ and reduced stiffness
		$L_p^{*}=L/b$. The solid boundary between globule and toroid bends toward higher
		$\epsilon^{*}$ at low stiffness, reflecting the persistence of isotropic globules
		at strong attraction and weak bending rigidity.}
	\label{fig:phase-diagram}
	
\end{figure}

%
%
%==================================
\section{Representative Ranges of $L_p^{*}$ and $\epsilon^{*}$ for Real Polymer Systems}
%======================================
Although the phase diagram in Fig.~\ref{fig:phase-diagram} is schematic, it is useful
to indicate realistic numerical ranges of the reduced persistence length
$L_p^{*}=L_p/b$ and reduced attraction strength $\epsilon^{*}=\epsilon/k_BT$ for
different classes of polymeric systems. These estimates provide empirical context
for the theoretical boundaries and clarify which regions of parameter space are
experimentally accessible.

\paragraph{Flexible synthetic polymers.}
Common synthetic polymers (polystyrene, PMMA, polyethylene oxide) typically have
$L_p \approx b$, giving
\[
L_p^{*} \sim 1\text{--}2.
\]
Solvent quality tunes the effective LJ attraction, with
\[
\epsilon^{*} \sim 0.8\text{--}1.2,
\]
spanning good to near-$\theta$ to weakly poor-solvent conditions. These systems lie
close to the coil--globule boundary.

\paragraph{Semiflexible LJ chains (simulation regime).}
Coarse-grained Lennard--Jones chains studied in simulations typically span
\[
L_p^{*} \sim 3\text{--}10,
\qquad
\epsilon^{*} \gtrsim 1.2,
\]
where globules, toroids, and short rods compete. This range corresponds precisely
to the regime explored by Srinivas and Bagchi.\cite{SrinivasBagchi2002}

\paragraph{Strongly semiflexible biopolymers (DNA).}
Double-stranded DNA has $L_p \approx 50\,\mathrm{nm}$; coarse-graining with
$b\sim1$--$2\,\mathrm{nm}$ gives
\[
L_p^{*} \sim 25\text{--}50.
\]
Salt-induced condensation corresponds to
\[
\epsilon^{*} \sim 1.3\text{--}1.8,
\]
placing DNA deep inside the ordered (toroid/rod) region of the phase diagram.

\paragraph{Very stiff polymers and rodlike filaments.}
Highly stiff conjugated polymers and biopolymers such as actin have
\[
L_p^{*} \sim 50\text{--}200,
\qquad
\epsilon^{*} \sim 1.0\text{--}2.0,
\]
and are firmly in the rod-dominated regime unless additional interactions stabilize
curved structures.

\paragraph{Summary.}
Across these systems,
\[
L_p^{*} \sim 1\text{--}200,
\qquad
\epsilon^{*} \sim 0.8\text{--}2.0,
\]
cover the experimentally relevant parameter space. These values confirm that the
schematic phase diagram captures realistic polymer behavior.

%=========================================================
\section{Competing Free-Energy Contributions and Morphological Transitions}
%=========================================================

The diverse morphologies adopted by a single polymer chain—coil, globule, toroid,
and rod—arise from competition among a small number of physical free-energy
contributions. Despite their geometric differences, the stability of these states
can be understood within a unified framework involving entropic elasticity, bulk
condensation energy, surface tension, and bending elasticity.

\subsection{Key free-energy contributions}

The dominant contributions are:

\begin{enumerate}
	\item \textbf{Entropic elasticity.}
	A swollen coil is stabilized by configurational entropy,
	\[
	F_{\mathrm{coil}} \sim k_BT
	\left( \frac{R^2}{Nb^2} + \frac{vN^2}{R^3} \right),
	\]
	with solvent quality encoded in the second virial coefficient $v$.
	
	\item \textbf{Bulk free energy.}
	All collapsed morphologies share the same bulk contribution,
	\[
	F_{\mathrm{bulk}} = N f_{\mathrm{bulk}}(\phi_0),
	\]
	reflecting the thermodynamics of the dense polymer phase.
	
	\item \textbf{Surface tension.}
	Interfaces with the solvent cost free energy,
	\[
	F_{\mathrm{surf}} = \gamma S,
	\]
	and distinguish compact (globule), elongated (rod), and curved (toroid) shapes.
	
	\item \textbf{Bending elasticity.}
	Chain stiffness penalizes curvature,
	\[
	F_{\mathrm{bend}} = \frac{\kappa}{2}\int ds\,R_{\mathrm{curv}}^{-2},
	\qquad
	\kappa = k_BT L_p.
	\]
	Toroids incur bending energy; rods largely avoid it.
\end{enumerate}

\subsection{Coil--globule transition}

Collapse occurs when entropic elasticity is overcome by bulk and surface terms:
\[
F_{\mathrm{coil}} = F_{\mathrm{glob}}
= N f_{\mathrm{bulk}}(\phi_0)
+ \tilde{A}_{\mathrm{glob}}\gamma a^2 N^{2/3}.
\]
This yields the finite-size crossover
\[
v_{\mathrm{c}}(N)
\sim
b^3
\left[
\frac{N^{4/5} f_{\mathrm{bulk}}(\phi_0)}{k_BT}
+
\tilde{A}_{\mathrm{glob}}
\frac{\gamma a^2}{k_BT}N^{3/5}
\right]^{5/2},
\]
showing that surface tension shifts the collapse to slightly positive $v$ for finite
chains.

\subsection{Globule--toroid transition}

Globules and toroids have identical bulk free energy; the competition is between
surface area and bending elasticity:
\[
F_{\mathrm{glob}}' \sim \gamma a^2 N^{2/3},
\qquad
F_{\mathrm{tor}}' \sim
(\kappa\gamma^2 a^3)^{2/5} (Nb)^{1/5} N^{1/5}.
\]
Equating these gives the crossover stiffness
\[
L_{p,\mathrm{c}}
\sim
\gamma^{1/2} a^2 b^{-1/2} N^{4/15}.
\]
Thus toroids are stabilized by increasing stiffness and surface tension.

\subsection{Toroid--rod transition}

Toroids reduce surface area but require curvature, whereas rods eliminate curvature
at the cost of end caps:
\[
F_{\mathrm{tor}}' \sim
(\kappa\gamma^2 a^3)^{2/5} (Nb)^{2/5},
\qquad
F_{\mathrm{rod}}' \sim \gamma a^2 N^{2/3}.
\]
The crossover occurs at
\[
L_{p,\mathrm{c}}
\sim
\gamma^{-1/2} a^{1/2} b^{-1} N^{4/15},
\]
showing that very stiff chains favor rods, while moderate stiffness favors toroids.

\subsection{Unified physical picture}

The morphology sequence follows directly from competing free-energy balances:

\begin{itemize}
	\item \textbf{Coil $\rightarrow$ Globule:} entropy vs.\ bulk and surface energy.
	\item \textbf{Globule $\rightarrow$ Toroid:} surface-area reduction vs.\ bending.
	\item \textbf{Toroid $\rightarrow$ Rod:} bending elasticity vs.\ end-cap cost.
\end{itemize}

The resulting $(L_p^{*},\epsilon^{*})$ phase diagram captures the observed
progression from flexible coils to globules, toroids, and rods. Helical structures
do not appear because the present free-energy functional contains no term favoring
torsion or chirality; in an isotropic wormlike-chain model, helices incur additional
elastic cost without compensating surface-energy gain. Stabilization of helices
therefore requires additional microscopic ingredients such as chiral or
directional interactions.\cite{GrosbergKhokhlov1994}

%======================================
\section{Triple-point estimate and relation to previous work}

The free-energy framework developed in this work allows one to place the
globule, toroidal, and rodlike morphologies of a semiflexible polymer on a
common footing and to compare their thermodynamic stability under identical
conditions. A particularly interesting consequence of this unified description
is the possibility of a junction in the morphology diagram at which the free
energies of these three distinct condensed states become comparable.

Within the present variational approach, the bulk free-energy contribution is
identical for all dense morphologies and cancels when comparing different
collapsed states. The relative stability is therefore controlled by subleading
terms arising from surface tension, bending elasticity, and orientational
ordering. Balancing these contributions yields an estimate of a point in the
$(\epsilon^{*},L_p^{*})$ plane at which the free energies of the globule, toroid,
and rodlike states coincide. For a long chain ($N=10^6$) and a reduced
persistence length $L_p^{*}\sim10$, appropriate for a stiff polymer, this
condition is found to occur at a reduced Lennard--Jones attraction strength
$\epsilon^{*}$ of order unity, well within the poor-solvent regime.

The physical ingredients underlying this balance are closely related to those
identified in earlier studies of stiff polymer condensation, most notably in
the seminal work of Odijk and collaborators, who emphasized the competition
between bending elasticity and surface tension in stabilizing toroidal and
rodlike condensates. The present analysis extends these ideas by embedding them
in a coarse-grained field-theoretic free-energy functional that also accounts
for density variations, nematic ordering, and their mutual coupling. This
unified framework naturally leads to a morphology diagram in which multiple
collapsed states appear as competing free-energy minima.

It is important to emphasize that the junction identified here should not be
interpreted as a sharp thermodynamic triple point in the strict sense. Finite
chain length, thermal fluctuations, and fluctuation-induced renormalization of
interfacial and ordering contributions are expected to broaden the vicinity of
this degeneracy into a narrow crossover region. Nevertheless, the estimated
location of this junction provides a concrete and experimentally testable
prediction of the theory, and offers a useful organizing principle for
interpreting simulations and experiments on semiflexible polymers in poor
solvents.

\section{Conclusions}

We have formulated a simple field-theoretic free-energy description of a semiflexible Lennard--Jones polymer that unifies the classical coil--globule transition with the emergence of ordered rod-like and toroidal collapsed states. By treating the dense polymer as a nematic melt with surface tension, and by adopting geometric ansatzes for the shapes, we derived analytic expressions for the free energies and crossover conditions in terms of effective interaction and energy parameters $(v,w,\gamma,L_p)$. The treatment is at the eman-filed level but we do consider role of fluctuations.

The resulting phase diagram in the $(L_p,\epsilon)$ plane is consistent with both earlier theoretical treatments of semiflexible polymer condensation and with the simulation results of Srinivas and Bagchi on LJ chains, as well as the extensive experimental literature on DNA toroids and rods. 

An example of how fluctuations can alter the phase boundary is provided by globule-rod transition where density fluctuation can induce nematic orser. The coupling between the two order parameters density phi and nematic order Q is also governed by the respective force constants. This adds interesting physics which deserved further studies.

The present framework can be systematically refined by including electrostatics, sequence heterogeneity, torsional rigidity, and coupling to external fields, and thus provides a useful starting point for future theoretical and computational studies of polymer condensation.

\appendix
In the appendices that follow, we discuss certain technical aspects that we glossed over or just stated in the main body of the text.%==================================================
\section{Orientational order and Landau--de~Gennes free energy}
\label{app:LDG}
%==================================================

Here we summarize the Landau--de~Gennes description of orientational order used in
the main text to characterize ordering within dense polymer globules.

\subsection{Nematic order parameter}

The local orientational order is described by the symmetric, traceless tensor
\begin{equation}
Q_{ij}(\mathbf{r})
=
\left\langle
u_i u_j - \frac{1}{3}\delta_{ij}
\right\rangle_{\mathbf{r}},
\label{eq:A1}
\end{equation}
where $\mathbf{u}$ is the local segment orientation and the average is taken over
segments within a coarse-graining volume around $\mathbf{r}$.
The isotropic state corresponds to $\mathbf{Q}=0$, while $\mathbf{Q}\neq 0$
signals nematic order.

For uniaxial order one may write
\begin{equation}
Q_{ij} = S \left( n_i n_j - \frac{1}{3}\delta_{ij} \right),
\label{eq:A2}
\end{equation}
where $\mathbf{n}$ is the director and $S$ the scalar order parameter.

\subsection{Rotational invariants}

The free-energy density must be rotationally invariant and is therefore
constructed from scalar invariants of $\mathbf{Q}$, notably
\begin{align}
\mathrm{Tr}\,\mathbf{Q}^2 &= \sum_{ij} Q_{ij}^2, \label{eq:A3} \\
\mathrm{Tr}\,\mathbf{Q}^3 &= \sum_{ijk} Q_{ij}Q_{jk}Q_{ki}. \label{eq:A4}
\end{align}
The quadratic invariant measures the magnitude of order, while the cubic
invariant distinguishes prolate and oblate nematic states and allows for a
first-order isotropic--nematic transition.

\subsection{Landau--de~Gennes free-energy density}

The Landau--de~Gennes free-energy density is written as
\begin{equation}
f_{\mathrm{nem}}(\phi,\mathbf{Q}) =
\frac{1}{2} A(\phi)\,\mathrm{Tr}\,\mathbf{Q}^2
- \frac{1}{3} B\,\mathrm{Tr}\,\mathbf{Q}^3
+ \frac{1}{4} C\,\left(\mathrm{Tr}\,\mathbf{Q}^2\right)^2
+ \frac{K}{2}(\nabla\mathbf{Q})^2,
\label{eq:A5}
\end{equation}
where $K$ is an elastic constant associated with spatial distortions of the
director field.

The density dependence enters through
\begin{equation}
A(\phi) = a_0(\phi-\phi^{*}),
\label{eq:A6}
\end{equation}
so that orientational order becomes favorable when the globule density exceeds
the threshold $\phi^{*}$. This mechanism underlies the emergence of nematic order
in sufficiently dense globules.\cite{KhokhlovSemenov1981,Odijk1986}

The density-gradient term
\begin{equation}
f_{\mathrm{dens,grad}} = \frac{\kappa_{\mathrm{el}}}{2}(\nabla\phi)^2
\label{eq:A7}
\end{equation}
generates an effective surface tension $\gamma$ for the globule, which enters the
shape-dependent free energies discussed in the main text.

%==================================================
\section{Density fluctuations and the Ornstein--Zernike propagator}
\label{app:OZ}
%==================================================

We derive here the density--density correlation function used in the main text.

\subsection{Gaussian theory of density fluctuations}

We consider small fluctuations about the uniform dense state,
\begin{equation}
\phi(\mathbf{r}) = \phi_0 + \delta\phi(\mathbf{r}),
\label{eq:B1}
\end{equation}
and expand the free-energy functional to quadratic order,
\begin{equation}
\mathcal{F}^{(2)}_\phi
=
\frac{1}{2}\int d^3 r
\left[
r_\phi (\delta\phi)^2
+
\kappa_{\mathrm{el}}(\nabla\delta\phi)^2
\right],
\label{eq:B2}
\end{equation}
where
\begin{equation}
r_\phi =
\left.
\frac{d^2 f_{\mathrm{bulk}}(\phi)}{d\phi^2}
\right|_{\phi=\phi_0}
\label{eq:B3}
\end{equation}
is the inverse compressibility of the dense phase.

\subsection{Ornstein--Zernike form}

In Fourier space the quadratic free energy diagonalizes,
\begin{equation}
\mathcal{F}^{(2)}_\phi
=
\frac{1}{2}\int \frac{d^3 q}{(2\pi)^3}
\left(
r_\phi + \kappa_{\mathrm{el}} q^2
\right)
|\delta\phi(\mathbf{q})|^2.
\label{eq:B4}
\end{equation}
The density propagator follows immediately:
\begin{equation}
G_\phi(q)
=
\langle
\delta\phi(\mathbf{q})\delta\phi(-\mathbf{q})
\rangle
=
\frac{1}{r_\phi + \kappa_{\mathrm{el}} q^2}
=
\frac{1}{\kappa_{\mathrm{el}}}
\frac{1}{q^2 + \xi_\phi^{-2}},
\label{eq:B5}
\end{equation}
where
\begin{equation}
\xi_\phi = \sqrt{\frac{\kappa_{\mathrm{el}}}{r_\phi}}
\label{eq:B6}
\end{equation}
is the density correlation length.

\subsection{Real-space correlations}

Fourier transformation yields the real-space correlation function
\begin{equation}
G_\phi(r)
=
\frac{1}{4\pi\kappa_{\mathrm{el}}}
\frac{e^{-r/\xi_\phi}}{r},
\label{eq:B7}
\end{equation}
showing that density correlations are short-ranged with characteristic decay
length $\xi_\phi$.

In the main text, integrating out $\delta\phi$ generates effective interactions
between orientational fluctuations mediated by $G_\phi$, with range controlled
by $\xi_\phi$.

\section* {Acknowledgment}

This work was supported partly by a India National Science Chair Professorship grant (from DST-Serb) to B.B.
%
%====================================

%
\end{document}